**Intraband versus interband scattering rate effects in neutron irradiated MgB$_2$**

M. Putti, P. Brotto, M. Monni and E. Galleani d'Agliano

*CNR-INFM-LAMIA and Dipartimento di Fisica, Università di Genova, Via Dodecaneso 33, 16146 Genova, Italy*

A. Sanna and S. Massidda

*CNR-INFM-SLACS and Department of Physics University of Cagliari, Cittadella Universitaria, I-09124 Monserrato (CA), Italy*

**Abstract**

One of the most important predictions of the two-gap theory of superconductivity concerns the role of interband scattering (IBS) by impurities. IBS is expected to decrease the critical temperature, $T_c$, of MgB$_2$ to a saturation value of about 20 K, where the two gaps merge to a single one. These predictions have been partially contradicted by experiments. In fact, $T_c$ does not saturate in irradiated samples, but decreases linearly with residual resistivity and the merging of the gaps has been observed at a much lower $T_c$ (11 K). In this paper we argue that, while at low level of disorder IBS is the leading mechanism that suppresses superconductivity, at higher disorder the experimental results can only be understood if the smearing of the density of states due to intraband electron lifetime effects is considered.

After the observation of two superconducting energy gaps in $MgB_2$ which open in different sheets of Fermi surface,[1,2] many investigations have been carried out to understand features and consequences of this peculiar phenomenon. Back in 1959 Matthias, Suhl and Walker[3] pointed out that, within the BCS theory, a higher transition temperature, $T_c$, should subsist if more variational degrees of freedom are provided, e.g. by allowing different order parameters in different bands. This is expected only for very clean samples because interband scattering (IBS) with impurities should suppress superconductivity in almost the same way, as scattering with magnetic ones does in a one-band superconductor.[4] When IBS rates become comparable with the relevant phonon frequency a complete isotropization over the whole Fermi surface is expected; the two gaps merge into one and $T_c$ drops to the isotropic value of about 20-25 K.[5,6,7] However, in $MgB_2$, the IBS rate is quite small even in rather dirty samples, due to the different symmetry of σ and π electronic states,[8] and it is believed that two distinct gaps are observable because of this fortunate coincidence. To verify these predictions several efforts have been done to introduce defects systematically by substitutions and by irradiation. Unfortunately, substitutional defects introduce charge doping, which complicates the understanding of the role of IBS itself. In this respect, irradiation with neutrons [9,10,11,12,13] and alpha particles[14,15] is very appealing since it produces homogeneous defect structures, without introducing charge doping. With increasing irradiation resistivity increases monotonously and $T_c$ decreases correspondingly; in heavily irradiated samples superconductivity is completely suppressed. By plotting $T_c$, as a function of the residual resistivity, $\rho_0$, a linear relation is found[13,14] without any sign of $T_c$ saturation around 20 K. A similar linear behaviour was observed in conventional superconductors like amorphous transition metals and damaged A15 superconductors,[16] which suggests that a common approach could be exploited.

The role of IBS has been mainly investigated by studying the evolution of the energy gaps, $\Delta_\sigma$ and $\Delta_\pi$, with disorder (Ref.[17] and references therein). While $\Delta_\sigma$ decreases linearly with $T_c$, $\Delta_\pi$ in weakly disordered samples increases in agreement with theoretical predictions;[18] however, with a further increase in disorder, $\Delta_\pi$ decreases and the merging of the gaps has finally been observed in neutron irradiated polycrystals at a critical temperature (11 K) much lower than the one predicted for isotropic MgB$_2$. These results prove that the IBS is able to drive from two- to single-gap superconductivity, but suggest that in irradiated samples other mechanisms cooperate to the suppression of T$_c$.

In this letter we show that a full understanding of the T$_c$ behavior in irradiated MgB$_2$ can be provided by introducing intraband electron lifetime effects in analogy with A15 superconductors. For these materials, the degradation of superconductivity induced by disorder was explained through the smearing of the density of states (DOS) produced by disorder. The model proposed by Testardi and Mattheiss[19] assumed that the smearing is due to the mixing of electron states in an energy region comparable to the inverse electron lifetime. In this letter we discuss the experimental behaviour of T$_c$ in irradiated MgB$_2$ samples within a generalization of Testardi-Mattheiss model to the multi-band case. The suppression of T$_c$ will be estimated in an anisotropic and an isotropic case and compared to experimental results.

The electron lifetime model proposed in Ref. 19 assumes that the defects just broaden the DOS via the electron relaxation rate $\Gamma$, which increases with disorder. If $N(E,\Gamma=0)$ is the DOS of a perfectly ordered material, disorder effects can be taken into account with the convolution:

$$N(E,\Gamma) = \int S(E,E',\Gamma) N(E',\Gamma=0) dE' \qquad (1)$$

where $S(E,E',\Gamma)$ is a broadening function that depends on the electron relaxation rate. Since $N(E,\Gamma)$ changes with increasing $\Gamma$, the Fermi energy in the disordered system, $E_F(\Gamma)$, is determined with the constraint that the total number of states remains constant.[20] The exact form of the broadening function is not crucial when the broadening is large compared to any fine structure in N(E). Within a semiclassical approach the proper weighting function should be a Lorentzian; however, to avoid complication from the broad Lorentzian wings, Testardi and Mattheiss suggested a thermal broadening with $S = -\partial f/\partial E$, where $f$ is the Fermi-Dirac function with T replaced by $T_B=\Gamma/k_B$, where $\Gamma$ is simply related to the residual resistivity $\rho_0$ by:

$$\Gamma = \rho_0 \Omega_p^2(\Gamma)/4\pi \qquad (2)$$

In the Testardi-Mattheiss approach $\Omega_p(\Gamma)$ is the plasma frequency of the disordered material obtained by applying to $\Omega_p(E,\Gamma=0)$ the same convolution of eq. (1).[19, 20] In this way the effects due to lifetime broadening are approximately accounted for and, with some additional assumptions, $T_c$ can be calculated as a function of $\rho_0$. In A15 superconductors[19, 20] the electron-phonon coupling constant, $\lambda$, was assumed to be proportional to the DOS of disordered material, $\lambda(\Gamma) \propto N(\Gamma)$ and, using the McMillan equation with fixed values of the pseudopotential $\mu^*$ and of the average phonon frequency, the degradation of $T_c$ was reproduced without farther free parameters, as a function of the measured $\rho_0$. In this work we generalize this model and use it to rationalize the behaviour of irradiated $MgB_2$.

In a two-band superconductor the effect of broadening has to be considered separately on the partial DOS (PDOS), $N_\sigma$ and $N_\pi$. In fact, as long as the IBS rates ($\Gamma_{\sigma\pi}$, $\Gamma_{\pi\sigma}=\Gamma_{\sigma\pi}N_\sigma/N_\pi$) are negligible in comparison with intraband scattering rates ($\Gamma_\sigma$, $\Gamma_\pi$), $\Gamma_\sigma$ and $\Gamma_\pi$ should produce the broadening of $N_\sigma$ and $N_\pi$, respectively. The hypothesis $\Gamma_{\sigma\pi}, \Gamma_{\pi\sigma} \ll \Gamma_\sigma, \Gamma_\pi$, well verified in pure $MgB_2$ due to the different parity of $\sigma$ and $\pi$ electronic states, is plausible in irradiated samples as long as defects do not disrupt significantly the *sp²* bonding pattern of the crystal.

The PDOS[21] and the plasma frequencies of the clean $MgB_2$ have been computed within the local density approximation using the FLAPW method and the broadening was taken into account by the Fermi-Dirac function. $\Omega_p(\Gamma)$ comes out to be practically unaffected by disorder, varying at most by 1%, while $N_\sigma(\Gamma)$ and $N_\pi(\Gamma)$ present significant deviations.

Fig. 1 shows $N(\Gamma)/N(0)$ as a function of $\Gamma$, for the $\sigma$ and $\pi$ band of $MgB_2$; here $N(\Gamma) = N[E_F(\Gamma),\Gamma]$ and $N(0)=N(E_F,0)$ is the DOS at the Fermi level of the clean material. For comparison we plot $N(\Gamma)/N(0)$ calculated in ref. 20 with the same approach for $V_3Si$ and $Nb_3Sn$. With increasing $\Gamma$ up to 0.2 eV, $N_\sigma(\Gamma)$ and $N_\pi(\Gamma)$ remain quite constant and then decrease; $N_\sigma(\Gamma)$, which is more reduced than $N_\pi(\Gamma)$, is about 80% of $N_\sigma(0)$ for $\Gamma=0.7$ eV. It is worth noting that a comparable reduction of DOS (0.75%) was estimated from NMR measurements in a strongly irradiated MgB$_2$ sample ($T_c$= 7 K).[22]

The behaviour of A15 is very different: $N(\Gamma)$ sharply decreases at low $\Gamma$ and then tends to saturate, being nearly one half of $N(0)$ for $\Gamma=0.4$ eV. These differences can be easily understood by looking at the DOS of the clean materials. A15 superconductors present a peak close to $E_F$[23] which quickly reduces by mixing the states. On the other hand, the DOS of MgB$_2$ is rather flat around $E_F$;[24] however, $N_\sigma$ becomes zero at 0.7 eV and when the smearing involves states at such a distance from $E_F$, $N_\sigma(\Gamma)$ decreases.

In the presence of multiple gaps, $T_c$ depends on the IBS rates. For negligible $\Gamma_{\sigma\pi}$, $T_c \equiv T_c^{(0)}$ is given, in the two-square-well model, by the multiband analog of the Allen-Mitrovic equation:[25,26]

$$\psi\left(\frac{\vartheta}{2\pi T_c^{(0)}}+1/2\right)-\psi(1/2)=(\Lambda^{(0)})^{-1} \qquad (3)$$

where $\psi(z)$ is the digamma function and $k_B\vartheta = \hbar\omega_c$ represents a characteristic cutoff phonon energy. $\Lambda^{(0)}$ is the largest eigenvalue of the matrix $\Lambda^{(0)}{}_{ij} = \dfrac{\lambda_{ij}-\mu_{ij}^*}{1+\sum_n \lambda_{in}}$ (i,j=$\sigma$,$\pi$), where $\lambda_{ij}$ is the asymmetric pairing interaction matrix and $\mu_{ij}^*$ is the Coulomb pseudopotential matrix. We notice that for $\vartheta \cong 600\text{-}700$ K and $T_c^{(0)} = 39$ K we have $\vartheta/2\pi T_c^{(0)}+1/2 \cong 3$, so that we cannot take the usual asymptotic expression $\psi(z) \sim \ln z$.

In the opposite limit, the IBS is larger than the relevant phonon frequency ($\Gamma_{\sigma\pi}\geq\omega_c$) and a full isotropization of all Fermi surfaces occurs. In this case only one gap is present and extending the analysis previously done in the pure BCS case we obtain the following expression for $T_c \equiv T_c^{(\infty)}$:

$$\psi\left(\frac{\vartheta}{2\pi T_c^{(\infty)}}+1/2\right)-\psi(1/2)=(\Lambda^{(\infty)})^{-1} \qquad (4)$$

where $\Lambda^{(\infty)} = \frac{1}{N_T^*} \left( \sum_{i,j=\sigma,\pi} N_i^* \Lambda_{ij}^{(0)} \right)$ and $N_i^* = N_i \left( 1 + \sum_{n=\sigma,\pi} \lambda_{in} \right)$ is the mass renormalized DOS of the i-band, with $N_T^* = \sum_{i=\sigma,\pi} N_i^*$.

To treat the general case, between $T_c^{(0)}$ and $T_c^{(\infty)}$, we have derived a generalization of eq. (3) which includes IBS. This equation takes the form: [27]

$$\psi\left(\frac{\vartheta}{2\pi T_c} + 1\right) - \psi(1/2) = \Lambda^{-1} \tag{5}$$

where, now, $\Lambda$ is the smallest eigenvalue of the matrix $(\Lambda^{-1})_{ij} = (\Lambda^{(0)})^{-1}_{ij} + \begin{vmatrix} n_j^* & -n_j^* \\ -n_i^* & n_i^* \end{vmatrix} y(T_c)$ with

$n_i^* = \frac{1}{N_T^*} N_i \left(1 + \sum_{n=\sigma,\pi} \lambda_{in}\right)$ and $y(T_c) = \psi(\frac{\vartheta}{2\pi T_c} + \frac{1}{2}) - \psi(\frac{1}{2}) - \psi\left(\frac{k_B \vartheta + \overline{\Gamma}}{2\pi k_B T_c} + \frac{1}{2}\right) + \psi\left(\frac{\overline{\Gamma}}{2\pi k_B T_c} + \frac{1}{2}\right)$.

Here $\overline{\Gamma}$ is proportional to $\Gamma_{\sigma\pi}$: $\overline{\Gamma} = \gamma \left( \frac{N_\sigma}{(1 + \sum_n \lambda_{\pi n})} + \frac{N_\pi}{(1 + \sum_n \lambda_{\sigma n})} \right)$, with $\gamma = \frac{\Gamma_{\sigma\pi}}{N_\pi} = \frac{\Gamma_{\pi\sigma}}{N_\sigma}$.

In order to take into account the effects of intraband scattering, the dependence of the matrices $\lambda_{ij}$ and $\mu_{ij}^*$ on disorder has to be considered. The pairing interaction matrix elements are proportional to the PDOS, $\lambda_{ij} = V_{ij} N_j$, where $V_{ij} = V_{ji}$ is the symmetric pairing potential matrix. Assuming $V_{ij}$ independent of $\Gamma$ we have:

$$\lambda(\Gamma) = \begin{vmatrix} \lambda_{\sigma\sigma} & \lambda_{\sigma\pi} \\ \lambda_{\pi\sigma} & \lambda_{\pi\pi} \end{vmatrix} = \begin{vmatrix} \lambda_{\sigma\sigma}(0)\frac{N_\sigma(\Gamma)}{N_\sigma(0)} & \lambda_{\sigma\pi}(0)\frac{N_\pi(\Gamma)}{N_\pi(0)} \\ \lambda_{\pi\sigma}(0)\frac{N_\sigma(\Gamma)}{N_\sigma(0)} & \lambda_{\pi\pi}(0)\frac{N_\pi(\Gamma)}{N_\pi(0)} \end{vmatrix} \tag{6}$$

where $\lambda_{\sigma\sigma}(0)=1.017$, $\lambda_{\sigma\pi}(0)=0.213$, $\lambda_{\pi\pi}(0)=0.448$ and $\lambda_{\pi\sigma}(0)=0.155$.[28] We also assume a scaling law of the pseudopotential matrix $\mu_{ij}^*$ with the PDOS; following ref.[29] we write:

$$\mu^*(\Gamma) = \begin{vmatrix} \mu^*_{\sigma\sigma} & \mu^*_{\sigma\pi} \\ \mu^*_{\pi\sigma} & \mu^*_{\pi\pi} \end{vmatrix} = \mu_0 N_T(\Gamma) \cdot \begin{vmatrix} \dfrac{2.23}{N_\sigma(\Gamma)} & \dfrac{1}{N_\sigma(\Gamma)} \\ \dfrac{1}{N_\pi(\Gamma)} & \dfrac{2.48}{N_\pi(\Gamma)} \end{vmatrix} \quad (7)$$

To check the reliability of our calculation, we calculate the Sommerfeld coefficient γ as a function of Γ and compare it with experimental values. γ(Γ) is given by:

$$\gamma(\Gamma) = \frac{2}{3}\pi^2 k_B^2 \left[ N_\sigma(\Gamma) \cdot \left(1 + \lambda_{\sigma\sigma}(\Gamma) + \lambda_{\sigma\pi}(\Gamma)\right) + N_\pi(\Gamma) \cdot \left(1 + \lambda_{\pi\pi}(\Gamma) + \lambda_{\pi\sigma}(\Gamma)\right) \right] \quad (8)$$

With $N_\sigma(0)=0.302$ states/eV cell and $N_\pi(0)=0.406$ states/eV cell from Eq. (8) we estimate $\gamma(0)=3.1$ mJ/mole K$^2$ to be compared with 3.0 mJ/mole K$^2$ and 2.5 mJ/mole K$^2$ [30] estimated from specific heat measurements. In Fig. 2, γ(Γ)/γ(0) is plotted as a function of Γ for MgB$_2$ and V$_3$Si: for V$_3$Si we assume λ(Γ)=λ(0)N(Γ)/N(0) with λ(0)=1.12 and N(Γ) from ref. 20. Reflecting the behaviour of the DOS, in MgB$_2$ γ(Γ)/γ(0) remains nearly constant for values of Γ less than 0.2 eV, and then slowly decreases. In V$_3$Si it sharply decreases at low Γ and then tends to saturate.

Experimental data can be plotted in Fig. 2 provided that the relaxation rate is known. For a single band metal, Γ can be simply estimated by the measured $\rho_0$ (see eq. (2)), while in the case of MgB$_2$ $\Gamma_\sigma$ and $\Gamma_\pi$ cannot be both evaluated. For the sake of simplicity in the following we assume equal intraband scattering rates in each band ($\Gamma_\sigma \sim \Gamma_\pi \equiv \Gamma$). This assumption, which is crude for substituted samples, is realistic when high degree of disorder is introduced homogeneously by irradiation. Experimental γ values estimated from specific heat in irradiated MgB$_2$ samples[17,30] and in V$_3$Si[31] are shown in Fig. 2 for comparison. The Γ values have been estimated for each sample from eq.(2) by keeping for MgB$_2$ $\Omega_p(\Gamma) \approx \Omega_p(0) = \sqrt{(\Omega_p(0)_\sigma)^2 + (\Omega_p(0)_\pi)^2} = 7.0$ eV (this value theoretically computed, has been recently confirmed by optical measurements[32]) and for V$_3$Si $\Omega_p(\Gamma) \approx \Omega_p(0) = 4.0$ eV.

As shown by Fig. 2, both in MgB$_2$ and V$_3$Si theoretical calculations reproduce quite well the experimental values, which indicates that the Testardi-Mattheiss model is capable of taking into account lifetime effects in materials which present very differently shaped DOS.

Now we are ready to discuss the T$_c$ vs $\Gamma$ behaviour. Experimental data obtained in neutron irradiated polycrystalline samples are shown in Fig. 3; on this set of samples the two-gap feature is present above 21 K (empty symbols) and the merging of the gaps was observed below 11 K (full symbols). Other comparable T$_c$ vs $\rho_0$ data reported in the literature[14,15] are not shown in Fig. 3 for the sake of clarity.

Continuous lines in Fig. 3 are the critical temperatures, $T_c^{(0)}$ and $T_c^{(\infty)}$, given by eqs (3) and (4) obtained by choosing $\omega_c$ = 53 meV ($\vartheta$ = 615 K) and $\mu_0$ = 0.050 (which implies $\mu_{\sigma\sigma}^*$=0.26, $\mu_{\sigma\pi}^*$=0.12, $\mu_{\pi\pi}^*$=0.22 and $\mu_{\pi\sigma}^*$=0.088) in good agreement with first principles calculations[5,28]. With the above choices of $\omega_c$ and $\mu_0$ we obtain $T_c^{(0)}(\Gamma=0)$=39.4 K and $T_c^{(\infty)}(\Gamma=0)$= 21.3, in agreement with other predictions,[5,6,7,33].

Starting from $\Gamma$=0, $T_c^{(0)}$ remains nearly constant for $\Gamma$ < 0.1 eV; then it decreases, reaching a value of about 18 K, for $\Gamma$ =0.7 eV ($\rho_0$~100 $\mu\Omega$cm). In this case, the IBS being completely neglected, $T_c^{(0)}$ decreases solely as a consequence of the reduction of the PDOSs, with increasing of intraband scattering. Also in the isotropic limit $T_c^{(\infty)}$ is nearly constant for $\Gamma$ < 0.1 eV values and it decreases as $\Gamma$ increases because of the PDOS reduction, reaching a value of about 8 K for $\Gamma$ =0.7 eV. Remarkably, the T$_c$ of the samples that present single-gap superconductivity (full symbols in Fig. 3) falls on $T_c^{(\infty)}$. On the other hand, the T$_c$ values of the samples that present two gaps (empty symbols) decrease roughly linearly with $\Gamma$ laying in between the two limiting curves for $T_c^{(0)}$ and $T_c^{(\infty)}$. In the range of weakly disordered samples ($\Gamma$≤0.1 eV), where the PDOSs are constant, this behaviour indicates that IBS is the main mechanism which suppresses T$_c$. When $\Gamma$ increases, also $\Gamma_{\sigma\pi}$ increases, but the two-gap feature survives, since the limit of strong IBS ($\Gamma_{\sigma\pi}<\omega_c$).is not yet reached .

The dotted line in Fig. 3 has been calculated by solving eq. (5) which takes into account both the effects of intraband scattering (PDOS reduction) and IBS. Assuming $\Gamma_{\sigma\pi}=\alpha\Gamma$ with $\alpha=0.059$, it reproduces pretty well the experimental values up to $\Gamma\sim0.3$ eV, while, for a further increase in disorder, the experimental data fall rather on $T_c^{(\infty)}$. Therefore, at low level of disorder the experimental data are well reproduced with $\Gamma_{\sigma\pi}$ increasing proportionally with $\Gamma$, the condition $\Gamma_{\sigma\pi}/\Gamma<<1$ being fulfilled; on the other hand above a certain level of disorder a faster increase of $\Gamma_{\sigma\pi}$ is expected because the $\pi$ and $\sigma$ states are loosing their symmetry. We point out that the samples showing a single gap must have $\Gamma_{\sigma\pi}\geq\omega_c\sim53$ meV with $\Gamma=0.5-0.6$ eV, so that $\Gamma_{\sigma\pi}/\Gamma<1$ remains true also at this level of disorder.

The details of our results depend on the assumption we made about the intraband scattering rates ($\Gamma_\sigma\sim\Gamma_\pi$). In weakly irradiated thin films the ratio $\Gamma_\sigma/\Gamma_\pi$ was estimated to be 1.5-2.[34] By varying $\Gamma_\sigma/\Gamma_\pi$ from 1 to 2, the values of $\omega_c$ and $\mu_0$, necessary to describe the experiments vary at most 10% ($\omega_c$ from 53 to 48 meV and $\mu_0$ from 0.050 to 0.046). The most affected parameter is $\alpha$ that varies from 0.059 to 0.035.

However, our main results are still confirmed: we rationalize the decrease of $T_c$ in irradiated samples, as well as the observation of the gap merging at 11 K rather than at the temperature predicted by IBS only (20-25 K). The latter point, which has been considered a puzzling problem so far, is naturally explained by considering that when the irradiation brings the system to the limit of strong IBS, the intraband scattering rate $\Gamma$ has obviously increased as well, reducing significantly the $\lambda_{ij}$. This mechanism could play a role also in doped $MgB_2$ samples, mainly when substitutions affect mostly intraband scattering of $\sigma$-bands, as it should occur in C-doped samples.

The analysis performed up to now, which provides a satisfactory explanation of the decreasing of $T_c$ for $\rho_0$ values less than 100 $\mu\Omega$cm, cannot be simply extended to the regime of extreme disorder.

First of all, in this regime experimental data are quite scattered because irradiation progressively reduces the connectivity between grains[13,14,15] and resistivity is hardly estimated; as a consequence, the resistivity at which $T_c$ is completely suppressed is not well defined. On the other hand, at such level of disorder some of our assumptions could fail. For instance, the electron mean free path of samples with $\rho_0 \sim 100$ μΩcm is of the order of the unit cell, and under these conditions one cannot neglect Anderson localization effects, which increase the effective Coulomb repulsion.[35]

We also point out that in our calculations we neglected changes in the phonon modes. Experimentally, the normal state specific heat increases weakly with irradiation and in the most damaged sample a reduction of the Debye temperature by 15% can be estimated, accompanied by an increase of the cell volume by about 1.7%. Raman spectroscopy of irradiated samples[36] shows that disorder causes the appearance of high frequency spectral structures, similar to those observed in doped $MgB_2$ samples that were considered as an evidence of violations of the selection rules rather than as a stiffening of the $E_{2g}$ mode. These results did not emphasize strong modifications of the phonon spectrum, but they cannot be ruled out, especially at high level of disorder. A detailed description of these effects, which will certainly affect to some extent the quantitative results, is beyond the purpose of the present investigation.

In conclusion, the comparison with A15 materials has suggested that also in irradiated $MgB_2$ electronic lifetime effects reduce the PDOS, mainly for the σ bands, producing a reduction of $T_c$ both in an anisotropic and an isotropic limit. The comparison with experiments suggests that at low level of disorder the pair breaking due to IBS plays the main role. However, when disorder increases and IBS progressively cancels the two-gap structure, the intraband scattering increases as well, reducing the PDOSs. As a result, the merging of the gaps is found at a critical temperature nearly one half of the value previously predicted, in excellent agreement with observation.

We acknowledge financial support by MIUR under the projects PRIN2004022024 and PON-CyberSar.

**Figure captions**

Figure 1. N(Γ)/N(0) as a function of relaxation rate Γ for the σ- and π-band of $MgB_2$, $V_3Si$ and $Nb_3Sn$ [20].

Figure 2. γ(Γ)/ γ(0) is plotted as a function of Γ for $MgB_2$ and $V_3Si$. Experimental data estimated from specific heat in irradiated $MgB_2$ and $V_3Si$ samples are reported for comparison.

Figure 3: $T_c$ values of neutron irradiated samples as a function of Γ: samples which present two gaps (empty symbols), samples which present one gap (full symbols). $T_c^{(0)}$ and $T_c^{(\infty)}$ given by eqs (3) and (4) as a function of Γ (continuous lines). $T_c$ values calculated by eq. (5) assuming $Γ_{σπ}=αΓ$ with α=0.059 (dotted line).

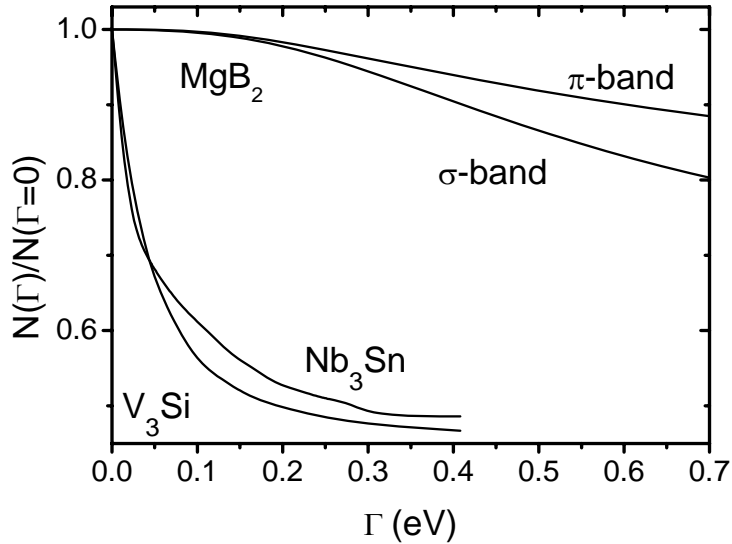

Figure 1

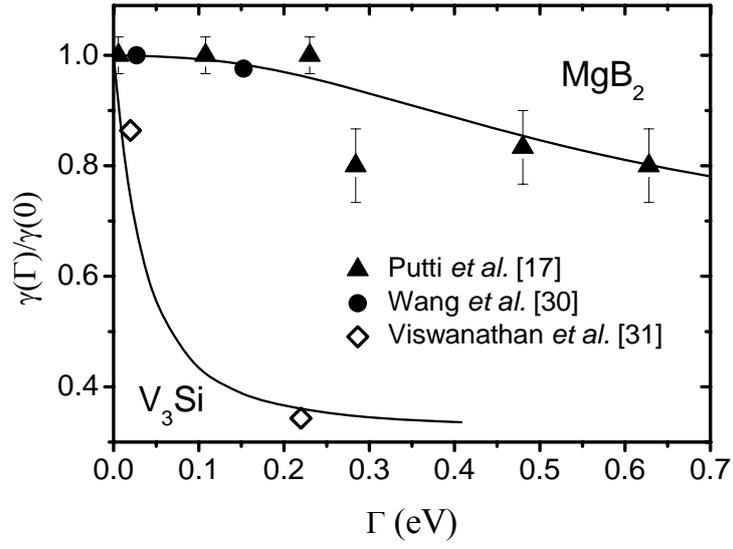

Figure 2

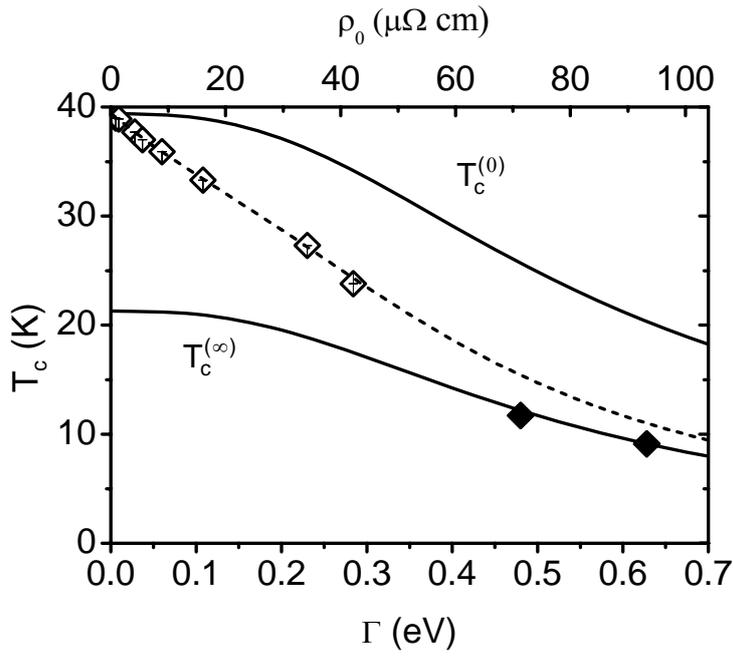

Figure 3